# Shared Renewable Allocation and Hydrogen Flexibility in Local Energy Markets: A Market Design Perspective

Pratik.Mochi, *Member, IEEE*, and Magnus Korpås, *Senior Member, IEEE*

***Abstract*— The integration of green hydrogen in local energy markets is often analyzed from a technical flexibility perspective, while the effect of market design rules remains less explored. This paper proposes a coordinated local electricity-hydrogen market framework in which hydrogen participation is regulated by explicit renewable access mechanisms. A mixed-integer linear programming model is developed to co-optimize electricity trading, battery operation, wind allocation and hydrogen production under centralized coordination. Six regulatory cases are examined including hydrogen supply options and access of local wind. Results are obtained for representative seasonal weeks for Norwegian energy community. Electrolyzer, when connected as rigid load, increases grid dependence, but also improves system cost when price-based participation is activated. Direct renewable access reduces grid imports, enhances wind allocation and introduces competition with households for energy distribution and system cost optimization. Furthermore, findings show that (i) hydrogen integration in local energy systems is essentially a market design problem and (ii) renewable access rules critically determine system behaviour, flexibility interactions and seasonal performance.**

## NOMENCLATURE

*Sets and Indices*

| | |
|---|---|
| $t \in \mathcal{T}$ | Time periods |
| $z \in \mathcal{Z}$ | Electricity end users |
| $w \in \mathcal{W}$ | Wind generation units |
| $e \in \mathcal{E}$ | Electrolyzers |

*Parameters*

| | |
|---|---|
| $P_{t,z}^{dem}$ | Electricity demand of end users (kW) |
| $P_{t,z}^{pv}$ | PV power generation of end users (kW) |
| $P_{t,w}^{wind}$ | Wind power generation (kW) |
| $P_{t,z}^{buy\ max}$ | Power import limit (kW) |
| $P_{t,z}^{sell\ max}$ | Power export limit (kW) |
| $B_{z}^{max}$ | Battery rated capacity (kW) |
| $P_{t,z}^{max,ch}$ | Battery charging limit (kW) |
| $P_{t,z}^{max,dch}$ | Battery discharging limit (kW) |
| $\eta_{z,ch}$ | Charging efficiency of battery |
| $\eta_{z,dch}$ | Discharging efficiency of battery |
| $₹_{t}^{grid,buy}$ | Grid purchase tariff (NOK/kWh) |
| $₹_{t}^{grid,sell}$ | Feed-in tariff (NOK/kWh) |
| $₹_{t}^{spot}$ | Market spot price (NOK/kWh) |
| $₹_{t}^{LEM}$ | Local Electricity Market price (NOK/kWh) |
| $₹_{t}^{grid,elz}$ | Grid buy tariff for electrolyzer (NOK/kWh) |
| $₹_{t}^{LEM,elz}$ | LEM price for electrolyzers (NOK/kWh) |
| $₹_{t}^{wind,z}$ | Wind power price to end users (NOK/kWh) |
| $₹_{t}^{wind,elz}$ | Wind power price to electrolyzers (NOK/kWh) |
| $₹_{t}^{wind,exp}$ | Wind power export price (NOK/kWh) |
| $D_{t}^{H2}$ | Total hydrogen demand (kg) |
| $S^{max,H2}$ | Hydrogen maximum storage capacity (kg) |
| $S^{min,H2}$ | Hydrogen minimum storage capacity (kg) |
| $S^{init}$ | Initial hydrogen storage level (kg) |
| $\eta^{ch,H2}$ | Hydrogen storage charging efficiency |
| $\eta^{dch,H2}$ | Hydrogen storage discharging efficiency |
| $P_{e}^{min,elz}$ | Minimum electrolyzer power (kW) |
| $P_{e}^{max,elz}$ | Maximum electrolyzer power (kW) |
| $\eta_{e}^{elz}$ | Electrolyzer efficiency |
| $LHV_{e}$ | Lower heat value conversion factor (kWh/kg) |
| $H_{e}^{max,pro}$ | Maximum hydrogen production rate (kg/h) |

*Decision Variables*

| | |
|---|---|
| $P_{t,z}^{grid,buy}$ | Power purchased from grid (kW) |
| $P_{t,z}^{grid,sell}$ | Power sold to grid (kW) |
| $P_{t,z}^{lem,buy}$ | Power purchased from LEM (kW) |
| $P_{t,z}^{lem,sell}$ | Power sold to LEM (kW) |
| $P_{t,z}^{import}$ | Total imported power (kW) |
| $P_{t,z}^{export}$ | Total exported power (kW) |
| $E_{t,z}^{bat}$ | Battery energy level (kWh) |
| $P_{t,z}^{ch}$ | Battery charging power (kW) |
| $P_{t,z}^{dch}$ | Battery discharging power (kW) |
| $P_{t,w,z}^{w\rightarrow z}$ | Wind power allocated to end users z (kW) |
| $P_{t,w,e}^{w\rightarrow elz}$ | Wind power allocated to electrolyzer e (kW) |
| $P_{t,e}^{w,pri}$ | Priority wind power to electrolyzer e (kW) |
| $P_{t,w}^{wind,exp}$ | Wind power exported to grid (kW) |
| $P_{t,e}^{elz}$ | Total electrolyzer input power (kW) |
| $P_{t,e}^{elz,grid}$ | Grid electricity to electrolyzer (kW) |
| $P_{t,e}^{elz,lem}$ | LEM electricity to electrolyzer (kW) |

This work was supported by The Research Council of Norway, grant number 326769.

Pratik Mochi and Magnus Korpås are with the Department of Electric Energy, Norwegian University of Science and Technology (NTNU), 7491 Trondheim, Norway (e-mail: pratikkumar.k.mochi@ntnu.no).

| | |
|---|---|
| $H_{t,e}^{prod}$ | Hydrogen production (kg) |
| $H_{t}^{sto}$ | Hydrogen storage level (kg) |
| $H_{t}^{rel}$ | Hydrogen supplied to demand (kg) |

## I. Introduction

LOCAL energy markets (LEMs) and community-based energy systems are gaining attention as a practical and inclusive way to manage high shares of renewable energy and provide flexibility at distribution level [1]. By allowing local trading among prosumers, LEMs can reduce grid stress, improve the use of local generation, and provide economic benefits to participating users [2]. At the same time, the rapid deployment of distributed wind and photovoltaic (PV) generation is changing how electricity systems operate, particularly in regions with strong renewable resources [3]. Alongside these developments, green hydrogen has emerged as a promising energy carrier for decarbonizing transport, industry and other energy intensive sectors [4]. Hydrogen produced via electrolysis using renewable electricity can act as a flexible energy sink by absorbing surplus generation and supporting renewable integration. However, hydrogen production often remains heavily dependent on grid electricity [5]. This can lead to more grid congestion and transmission losses, as well as higher electricity costs.

Local hydrogen production, supported by nearby renewable resources, offers a potential solution to these challenges. When electrolyzers are placed within local energy systems, they can respond to local generation and demand. It reduces reliance on centralized hydrogen supply chains and avoids transportation-related losses [6]. However, introducing hydrogen production at the local level also creates new interactions and challenges within the energy system. In particular, hydrogen electrolyzers behave as large, flexible electrical loads that may compete with existing flexibility options [7] such as household batteries, variable renewables and demand shifting. While batteries typically serve short-term balancing needs, electrolyzers can consume significant amounts of electricity over longer periods when equipped with sufficient hydrogen storage and/or distribution systems. Understanding how these different flexibility resources interact within a local market is therefore essential for designing effective and fair market mechanisms [8].

### A. Related Literature

Existing research on local energy markets has often focused on electricity-only systems, where households and prosumers trade electricity locally to reduce costs and improve renewable utilization [9]. These studies demonstrate that coordinated local trading can lower grid imports, smooth load profiles and increase the self-consumption of renewable energy. In parallel, some interesting work examines hydrogen integration into power systems, often at the transmission or system-wide level [10-12]. These studies highlight the role of electrolyzers as flexible loads that can respond to electricity prices and help integrate variable renewable energy. Several works also explore joint optimization of electricity and hydrogen systems, typically considering centralized planning or large-scale market configurations [13-14]. More recent studies combine local electricity and hydrogen within coordinated optimization frameworks, showing potential cost reductions and improved renewable usage when both energy carriers are considered together [15-18]. Market-based approaches, where hydrogen production responds to electricity prices, are commonly used to represent flexibility.

Despite these advances, important aspects remain underexplored at the local level. Most existing models do not explicitly consider how renewable generation is allocated among competing local users, such as households and electrolyzers. Competition between hydrogen production and prosumers for locally generated wind or PV power is often neglected or implicitly resolved through prices alone. Furthermore, market access rules such as priority access, shared access or capped participation are rarely modeled, even though such rules can strongly influence system outcomes. Local pricing mechanisms for electricity used in hydrogen production are also typically simplified or assumed to follow wholesale market signals.

### B. Research Gap and Problem Statement

When electrolyzers are operated as a fixed grid connected load or purely as a price responsive consumer, it is difficult to assess their full impact on the system [19]. In such cases, they will increase grid imports and possibly also operating cost without contributing to local renewable integration. On the other hand, when hydrogen production is modeled as responding only to local electricity prices, the physical availability of renewable resources and the resulting competition with other local users are not fully represented. Market design decisions are therefore important in determining how hydrogen interacts with the energy systems [20]. Rules governing access to local renewable generation would affect who can use renewable power, when hydrogen production takes place, and how the benefits of flexibility are shared among participants. Priority access for hydrogen, unrestricted competition with end users or explicit limits on renewable usage can all lead to very different outcomes in terms of system operation and cost distribution.

This research examines hydrogen integration in local energy markets that not only considers price signals but also market access and allocation rules. The main question under investigation is how different hydrogen participation mechanisms affect renewable usage, grid dependency, household flexibility and overall system performance across seasons.

### C. Contribution

The main contributions of this paper are as follows.

1. A local energy market framework is proposed, coordinated by a central aggregator that optimizes system operation on behalf of local participants.
2. A mixed-integer linear programming (MILP) model is developed to jointly optimize electricity trading, hydrogen production, storage operation, and renewable allocation.
3. Six different hydrogen integration cases are

introduced, representing different regulatory and market access mechanisms in the local energy market. The impact of these mechanisms on renewable allocation, grid imports, battery usage, and hydrogen production costs is quantitatively assessed.

4. Seasonal variations are considered using representative weeks to analyze differences in renewable availability and demand patterns.

## II. Market Framework for Local Electricity-hydrogen Trading

This section presents the proposed local energy market framework that defines the scope and boundaries of the optimization model developed in this paper. We have defined market participants, their interactions and the rules governing access to local renewable resources. This market-centric formulation allows the impacts of different hydrogen integration strategies to be assessed not only in terms of system costs, but also in terms of resource allocation and participation outcomes.

### *A. Market Participants & Roles*

The proposed local market consists of multiple interacting participants operating at the distribution level. Each of them plays a clearly defined role in the local energy system.

- End users represent residential, commercial or industrial buildings equipped with electricity demand, rooftop photovoltaic generation, and battery energy storage systems. End users can purchase electricity from the grid or the local energy market, consume locally generated renewable power, store energy in batteries, and export surplus electricity either to the grid or to other market participants through the local market.
- Wind generation units are modeled as local renewable energy sources connected at the distribution level. Wind power can be allocated to end users, used for local hydrogen production, or exported to the upstream grid, depending on market rules and system conditions.
- Electrolyzers represent local hydrogen production units that convert electricity into hydrogen. These electrolyzers may procure electricity from the grid, the local energy market, or directly from local wind generation, depending on the regulatory case under consideration. Electrolyzers act as large, controllable electricity consumers and introduce an additional level of flexibility into the local system.
- Hydrogen demand entities include hydrogen buses and ferries. Their hydrogen demand must be satisfied at all times, either from real-time hydrogen production or from hydrogen storage. The hydrogen demand from ferries is an inflexible demand while hydrogen demand from buses depends on factors like seasonal effect and weekend frequencies.
- The Distribution System Operator (DSO) provides access to the main electricity grid. The DSO supplies electricity at time-varying grid tariffs and remunerates exported electricity through feed-in tariffs. The DSO is not involved in local market clearing but serves as an external interface for electricity imports and exports.
- The Local Energy Market Aggregator (LEMA) acts as the central coordinating entity of the local market. The LEMA collects information from all participants and determines optimal system operation while respecting market access rules and technical constraints.

### *B. Electricity & Hydrogen Trading Logic*

Electricity trading within the local market occurs through multiple ways. End users can purchase electricity from the grid, trade electricity with other end users through the local energy market or consume locally generated wind and PV power. Surplus electricity can be exported either to the local market or to the upstream grid, subject to technical limits. Local wind generation plays a dual role in the system. Wind power may be allocated to meet the electricity demand of end users, charge batteries, supply electrolyzers for hydrogen production, or be exported to the grid. The allocation of wind power is explicitly modeled and constrained to ensure that total usage does not exceed available generation. Hydrogen trading follows a simpler logic. Hydrogen is produced locally by electrolyzers and stored in a hydrogen storage system. Stored hydrogen is then released to meet the demand of hydrogen buses and ferries. Hydrogen demand is assumed to be inflexible and must be fully satisfied in all time periods. This joint modeling of electricity and hydrogen flows allows the electrolyzer to act as a link between the two energy carriers. As a result, decisions in the electricity market directly affect hydrogen production patterns and vice versa.

### *C. Market Access Rules*

Hydrogen integration in a local energy market is not driven by prices alone. In practice, access to local renewables and market platforms is often governed by predefined rules. In this study, the electrolyzers do not automatically receive equal access to wind power or local market trades. Instead, their participation depends on specific access conditions. These conditions define whether hydrogen production is limited to grid electricity, allowed to trade in the local market, or permitted to use locally generated wind power. They also determine whether electrolyzers compete with prosumers on equal terms or receive priority. The detailed structure of these rules is not defined in this section. It is introduced through the six case studies presented in Section V.

### *D. Market Framework*

The market structure considered in this work is coordinated by a local energy market aggregator, as shown in Fig. 1. End users, wind generation, and the electrolyzer participate in a common local electricity market, while remaining connected to the distribution grid operated by the DSO. The DSO provides retail purchase tariffs, feed-in tariffs, and import/export limits. These external signals form the boundary conditions of the local market. Market participants submit their expected electricity demand, local generation, and hydrogen demand in advance. Based on this information, LEMA determines the optimal electricity exchanges among participants and the grid. The local market price is derived from the cost minimization framework and is applied to electricity traded between end users and the electrolyzer. Hydrogen demand from the ferry and bus is treated as an exogenous requirement and must be satisfied through local production. Communication links between LEMA, the DSO, and the participants allow coordinated scheduling. Physical energy flows and economic coordination are therefore separated: electricity and hydrogen follow the network shown in Fig. 1, while trading decisions and price formation are handled centrally by LEMA. The objective of the LEMA is to minimize the total system cost on behalf of all market participants. This system-level objective includes electricity purchase costs, revenues from electricity exports, and the electricity cost associated with hydrogen production. At the same time, the aggregator ensures that all electricity and hydrogen balance constraints are satisfied and that market access rules are respected.

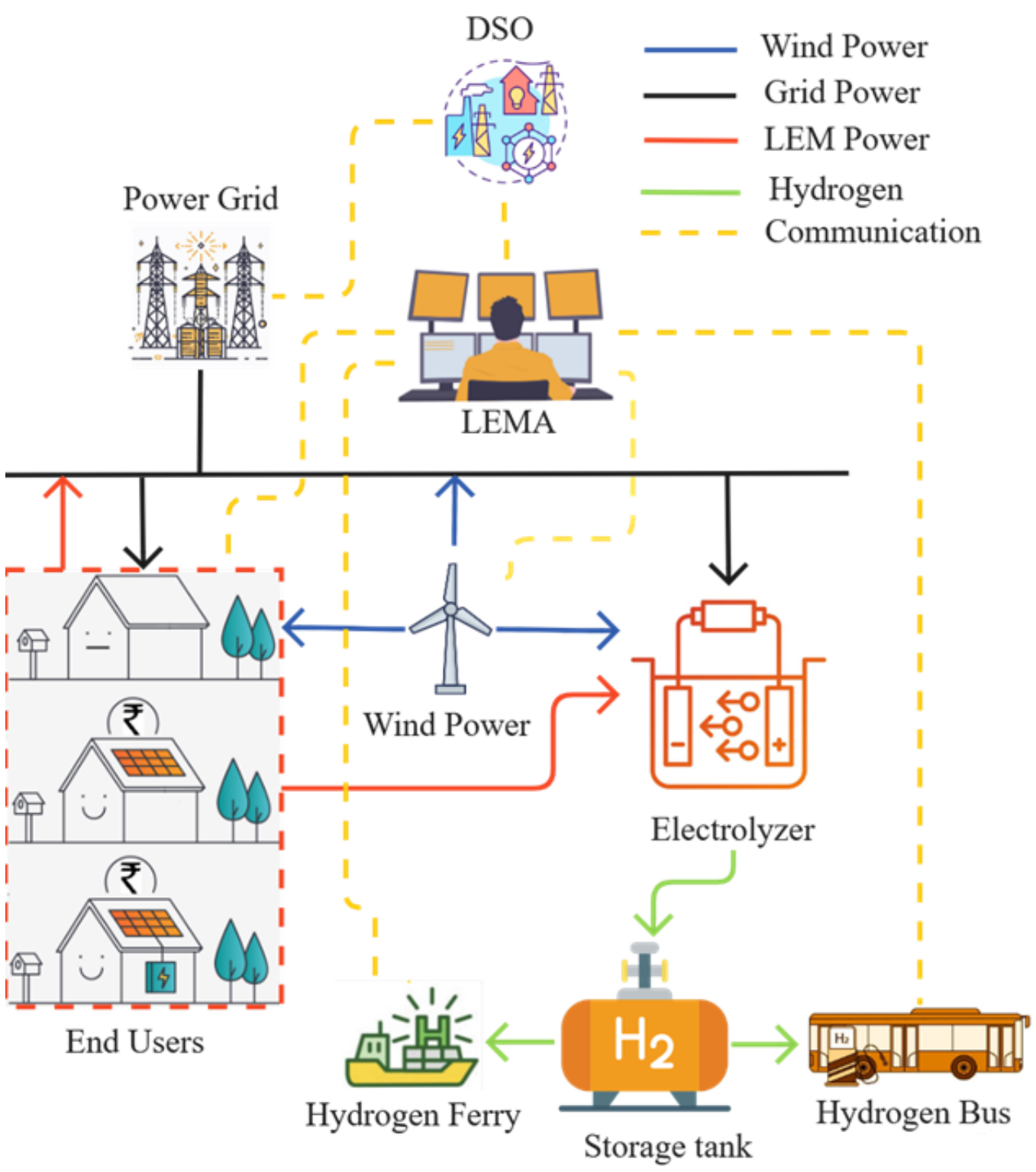

Fig. 1. LEM Overview

## III. System Description & Input Data

This section consists of brief discussion on physical system overview, time durations and seasonal consideration, and datset used to analyze proposed LEM.

### A. Physical System Overview

The physical structure of the modeled system is illustrated in Fig. 1. End users represent residential or commercial consumers equipped with local photovoltaic generation and, where applicable, battery storage. These users are connected to a common electricity bus that also receives power from wind generation units and the external grid. Electricity can flow between end users, wind generation, the electrolyzer, and the grid, subject to operational limits. The electrolyzer converts electrical energy into hydrogen. Its power input may originate from wind power, local electricity exchanges, or the grid. The produced hydrogen is stored in a dedicated storage tank before being supplied to final demand. In the present configuration, hydrogen demand is represented by a ferry and a bus, which draw hydrogen from the storage unit according to predefined schedules. All electricity interactions occur at a single node, where power balance is enforced at each time step. Hydrogen storage is modeled with capacity and efficiency constraints, ensuring continuity between production and consumption.

### B. Time Horizon & Resolution

The optimization model is formulated over a discrete-time horizon with an hourly resolution. To include seasonal variability without incurring excessive computational burden, four representative weeks are selected to represent winter, spring, summer, and late summer conditions. Each representative week consists of 168 hourly time steps, resulting in a total of 672 modeled hours across all seasons. The weeks are treated independently in the optimization, allowing seasonal differences in renewable availability, load patterns, and market prices to be analyzed. This approach is consistent with common practice in energy system studies and enables a clear comparison of system behavior for various operating conditions. The hourly resolution allows the model to describe short-term operational decisions such as battery charging and discharging, electrolyzer scheduling, and local market trading.

To include seasonal operational characteristics under Norwegian conditions, four representative weeks were selected from the 2024 dataset. The selected weeks correspond to deep winter (January 15-21), spring transition (April 8-14), summer peak solar availability (June 17-23) and late-summer conditions (August 19-25) as shown in table 1. Their symbolic representations are W1, W2, W3 and W4 respectively. These weeks present individual electricity price schedules, renewable generation profiles and load patterns. In winter, high demand combined with limited solar generation and higher electricity prices stresses hydrogen production and storage. Whereas summer conditions present higher solar generation, lower prices, and potential hydrogen surplus. Transitional seasons show mixed renewable dominance and variable price signals, providing a rigorous test for the proposed local energy market–based hydrogen production framework. This selection permits a comprehensive evaluation of market efficiency, renewable utilization, and hydrogen system flexibility through seasons.

TABLE I. Week Selection

| Week | Load | PV generation | Wind generation | Electricity prices |
|---|---|---|---|---|
| W1 | Highest | Lowest | High | Highest |
| W2 | Moderate | Rising | Moderate | Low |
| W3 | Lowest | Highest | Moderate | Low & volatile |
| W4 | Rising | Declining | Highest | Rising |

### C. Input Data Sources

The simulations are based on an open-source dataset representing a Norwegian local energy community [21], obtained from the Zenodo repository [22]. The system parameters are shown in table 2, figure 2 and figure 3.

TABLE II. SYSTEM PARAMETERS

| Parameter | Value | Parameter | Value |
|---|---|---|---|
| Z | 300 | $P_{t,z}^{pv}$ (kW) | 0-16.5 |
| W | 2 | $B_z^{max}$ (kW) | 0-17.2 |
| E | 2 | $\eta_{z,ch}, \eta_{z,dch}$ | 0.945-0.97 |
| T | 168 | $D_t^{H2}$ (kg) | 3268 (W1)<br>3087 (W2)<br>2671 (W3)<br>3084 (W4) |
| $P_e^{max,elz}$ (MW) | 1.2, 2.5 | | |
| LHV (kWh/kg) | 58, 52 | | |
| $\eta_e^{elz}$ | 0.73, 0.63 | | |
| $H_e^{max,pro}$ (kg/h) | 20.83, 45 | $\text{₹}_t^{grid,sell}$, $\text{₹}_t^{wind,z}$, $\text{₹}_t^{wind,elz}$ (NOK/kWh) | 0.789 (W1)<br>0.557(W2)<br>0.557(W3)<br>0.353 (W4) |
| $S^{max,H2}$ (kg) | 450 | | |
| $\eta^{ch,H2}$, $\eta^{dch,H2}$ | 0.98 | | |
| $P_{t,w}^{wind}$ (MW) | 0-1.3, 0-0.85 | | |

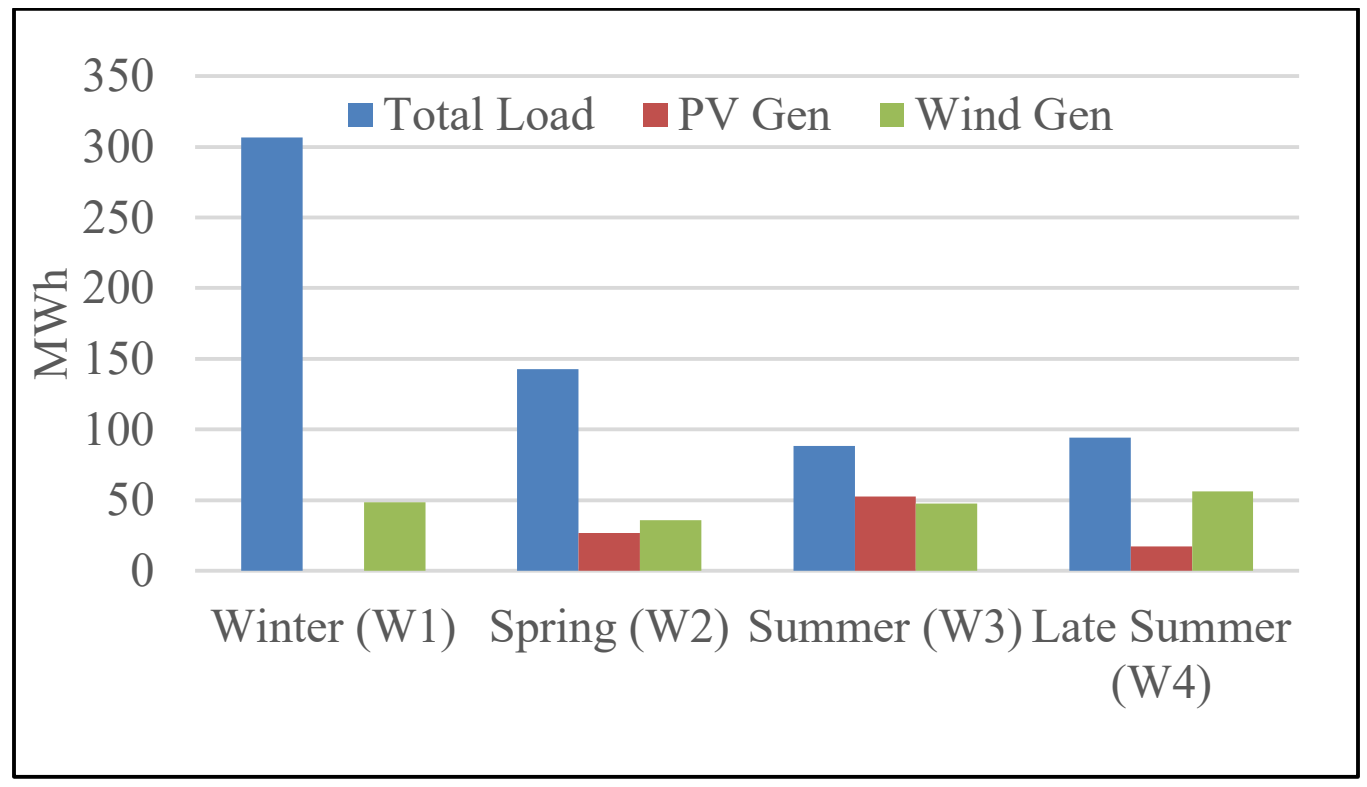

Fig. 2. Weekly generation & load of end users

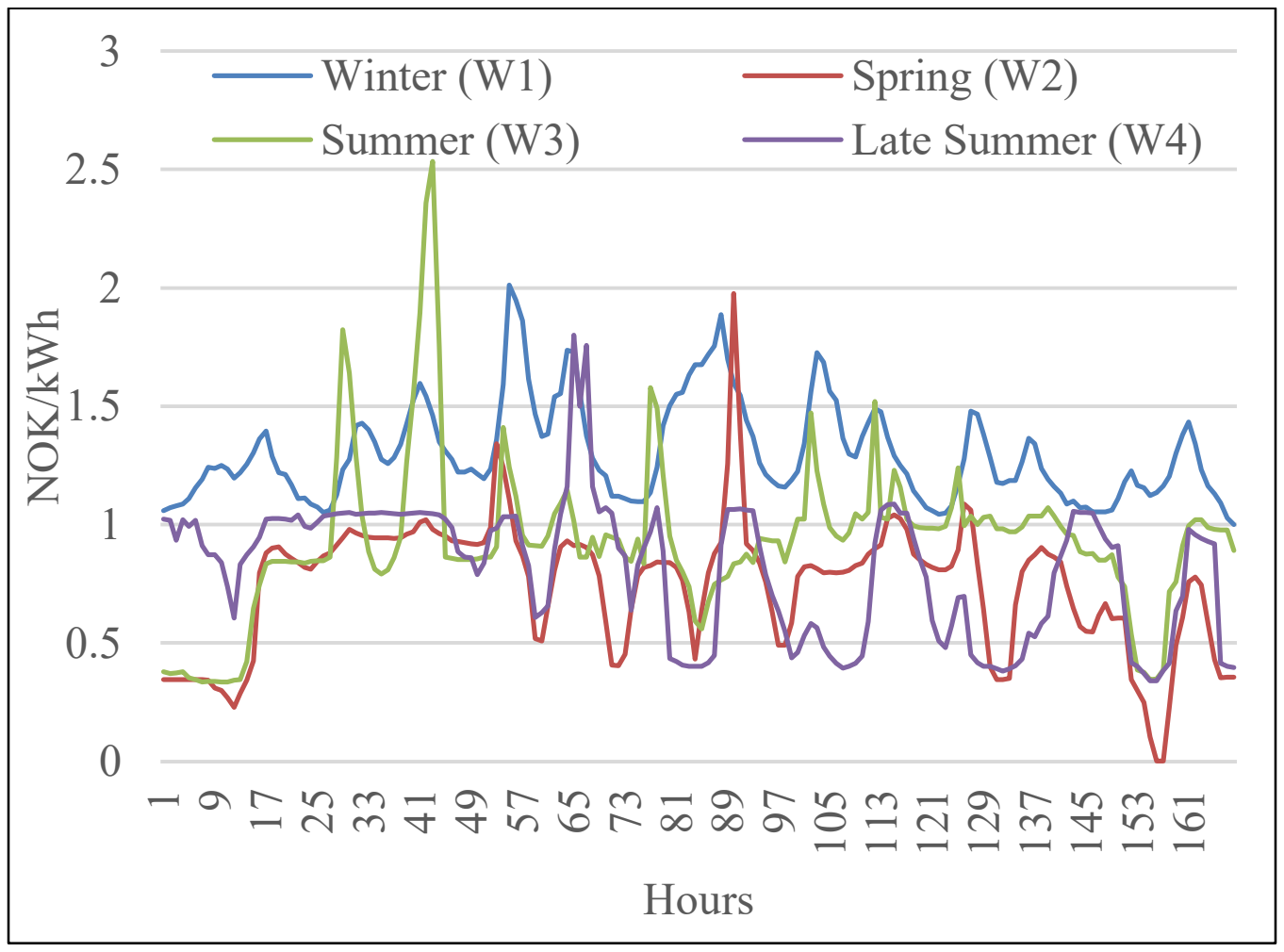

Fig. 3. Electricity price for end users & electrolyzer

## IV. OPTIMIZATION MODEL FORMULATION

This section presents the mathematical formulation of the proposed LEM model. The model is developed as a MILP problem and is solved from the perspective of the LEMA. The aggregator co-optimizes electricity trading, renewable allocation, storage operation, and hydrogen production while respecting physical limits and market access rules. The formulation is generic and flexible such that it allows different regulatory designs for hydrogen participation to be implemented through case-specific constraints.

### A. Objective Function

The objective function aggregates all relevant electricity and hydrogen related costs and revenues throughout participants and time periods.

$$\begin{aligned}\text{Min} \sum_{t=1}^{T} \Bigg( \sum_{z=1}^{Z} \Big( & \text{₹}_t^{grid,buy} P_{t,z}^{grid,buy} + \text{₹}_t^{LEM} P_{t,z}^{lem,buy} \\ & - \text{₹}_t^{grid,sell} P_{t,z}^{grid,sell} - \text{₹}_t^{LEM} P_{t,z}^{lem,sell} \Big) \\ & + \sum_{e=1}^{E} \Bigg( \text{₹}_t^{grid,elz} P_{t,e}^{elz,grid} \\ & + \text{₹}_t^{LEM,elz} P_{t,e}^{elz,lem} \\ & + \text{₹}_t^{wind,elz} \sum_{w=1}^{W} P_{t,w,e}^{w\to elz} \Bigg) \\ & + \text{₹}_t^{wind,z} \sum_{z=1}^{Z} P_{t,w,z}^{w\to z} \\ & - \sum_{w=1}^{W} \text{₹}_t^{wind,exp} P_{t,w}^{wind,exp} \Bigg)\end{aligned} \tag{1}$$

### B. Core Constraints

The equation 2 defines the power balance equation of each end-user. This constraint ensures that local demand, battery storage and power export are balanced by local generation, battery usage and power import, in each time period.

$$\begin{aligned} P_{t,z}^{grid,buy} + P_{t,z}^{lem,buy} + P_{t,z}^{dch} + P_{t,z}^{pv} + \sum_{z=1}^{Z} P_{t,w,z}^{w\to z} \\ = P_{t,z}^{dem} + P_{t,z}^{grid,sell} + P_{t,z}^{lem,sell} + P_{t,z}^{ch}\end{aligned} \tag{2}$$

The market clearing within LEM is constrained by equation 3, which balances internal LEM power.

$$\sum_{z=1}^{Z} P_{t,z}^{lem,buy} = \sum_{z=1}^{Z} P_{t,z}^{lem,sell} + P_{t,e}^{elz,lem} \tag{3}$$

Electricity imports and exports are bounded by grid capacity limits and local market limits. Binary variables impose mutually buying and selling behavior within each market in a given time period. Battery operation is governed by standard state-of-charge balance equations, eq. 4 and 5. Charging and discharging efficiencies are explicitly modeled, and binary constraints ensure that batteries cannot charge and discharge simultaneously.

$$E_{t,z}^{bat} = E_{t-1,z}^{bat} + P_{t,z}^{ch} \times \eta_{z,ch} - \frac{P_{t,z}^{dch}}{\eta_{z,dch}} \tag{4}$$

$$0 \le E_{t,z}^{bat} \le B_z^{max} \tag{5}$$

After the production of hydrogen, the storage and release of hydrogen are considered in the system and are calculated by equations (6-8). The electrolyzer power consumption is calculated by equation (9) provided the power limits and operational constraints (10-13).

$$D_t^{H2} = H_t^{rel} \tag{6}$$

$$H_1^{sto} = S^{init} + H_{1,e}^{prod} \times \eta^{ch,H2} - \frac{H_1^{rel}}{\eta^{dch,H2}} \tag{7}$$

$$H_t^{sto} = H_{t-1}^{sto} + H_{t,e}^{prod} \times \eta^{ch,H2} - \frac{H_t^{rel}}{\eta^{dch,H2}} \tag{8}$$

$$P_{t,e}^{elz} = \frac{H_{t,e}^{prod} \times LHV}{\eta_e^{elz}} \tag{9}$$

$$P_{t,e}^{elz} = P_{t,e}^{elz,grid} + P_{t,e}^{elz,lem} + \sum_{w=1}^{W} P_{t,w,e}^{w\to elz} \tag{10}$$

$$P_e^{min,elz} \le P_{t,e}^{elz} \le P_e^{max,elz} \tag{11}$$

$$S^{min,H2} \le H_t^{sto} \le S^{max,H2} \tag{12}$$

$$H_{t,e}^{prod} \le H_e^{max,pro} \tag{13}$$

The power transaction limits by end users with grid and LEM are constrained by equations (14-17).

$$0 \le P_{t,z}^{grid,buy} \le P_{t,z}^{buy\ max} \tag{14}$$

$$0 \le P_{t,z}^{grid,sell} \le P_{t,z}^{sell\ max} \tag{15}$$

$$0 \le P_{t,z}^{lem,buy} \le P_{t,z}^{buy\ max} \tag{16}$$

$$0 \le P_{t,z}^{lem,sell} \le P_{t,z}^{sell\ max} \tag{17}$$

Binary variables are used to implement logical market conditions such as exclusive access to grid or LEM electricity, limited simultaneous use of electricity sources, conditional activation of storage and electrolyzer operation. These constraints are essential for accurately representing discrete market participation decisions.

### C. *Wind Allocation Constraints*

Wind power is conserved and allocated through a set of allocation constraints that link generation to consumption and export. For each wind turbine and time period, total available wind generation must equal the sum of wind power allocated to prosumers, wind power allocated to electrolyzers, wind power exported to the grid.

$$P_{t,w}^{wind} = \sum_{z=1}^{Z} P_{t,w,z}^{w\to z} + \sum_{w=1}^{W} P_{t,w,e}^{w\to elz} + P_{t,w}^{wind,exp} \tag{18}$$

$$\sum_{z=1}^{Z} P_{t,w,z}^{w\to z} \le P_{t,z}^{dem} + P_{t,z}^{ch} \tag{19}$$

Additional constraints limit wind allocation to prosumers based on their instantaneous demand and battery charging capacity, preventing unrealistic overconsumption of renewable power. For electrolyzers, wind usage is bounded by electrolyzer capacity and market access rules defined by the specific case.

### D. *Calculation of local market prices*

LEM prices are constructed endogenously based on external grid prices and feed-in tariffs [16]. For each day, the minimum hourly spot price is identified, and the local electricity price is defined as the average of this minimum spot price and the applicable grid feed-in tariff. This approach ensures that local electricity prices lie between grid import and export prices, creating incentives for local trading while maintaining economic consistency.

$$₹_t^{LEM} = \frac{\min(₹_t^{spot}) + ₹_t^{grid,sell}}{2} \tag{20}$$

A similar approach is used to construct the electricity price faced by electrolyzers when purchasing electricity from the local market. The local hydrogen electricity price is defined on a daily basis and reflects the same underlying price signals as the prosumer-facing LEM price, while allowing for differentiated treatment of hydrogen production if required by the market design.

### E. Limitations

Several limitations define the scope of the present study. First, the model adopts an interpretation based on representative seasonal weeks. Uncertainty in renewable generation, electricity prices, and hydrogen demand is not explicitly modeled. Second, the local energy market is represented as a single-node system without detailed distribution network constraints. Line congestion, voltage limits, and spatial effects are not considered. Third, the optimization assumes centralized coordination by an aggregator and does not model strategic bidding behavior or decentralized market clearing mechanisms. Fourth, the analysis focuses on operational decisions under fixed infrastructure capacities. Investment planning for electrolyzers, renewable expansion, or storage sizing is not considered. These assumptions allow a clear comparison of hydrogen integration mechanisms but limit the direct generalization of quantitative results to more complex systems.

## V. Case Description & Methodology

To evaluate the hydrogen integration strategies in local electricity-hydrogen market, various market design and regulatory cases are defined. All cases are implemented on the same physical system and data set described in Section III and are solved using the optimization framework introduced in Section IV. The cases differ only in the market access rules governing hydrogen production, allowing a controlled comparison of how regulatory design choices affect system operations.

### A. *Case 0 – No Hydrogen*

Case 0 represents the reference scenario without hydrogen production and serves as the benchmark for all subsequent comparisons. In this case, the local energy market includes only electricity trading among residential prosumers, battery storage, local renewable generation, and grid interaction. Wind and PV generation are allocated exclusively to household demand, battery charging and grid export. No electrolyzers, hydrogen storage, or hydrogen demand are present in the system. The local energy market aggregator optimizes electricity dispatch and trading to minimize total system cost, subject to power balance, storage dynamics and grid limits. This case establishes the baseline behavior of the local electricity market and provides a lower-bound reference for system cost, renewable utilization and battery usage against which the impact of hydrogen integration is evaluated. To implement this case, all the hydrogen related parameters and equations are removed from the model.

*B. Case 1 – Grid-Only Hydrogen*

Case 1 introduces hydrogen production into the system but restricts electrolyzers to use electricity only from the main grid. Local market electricity and renewable generation are not accessible to hydrogen production. Hydrogen demand from buses and ferries must be met at all times through electrolyzers powered by grid electricity. While the electrolyzers are subject to operational constraints such as capacity limits and efficiency, they do not participate in the local electricity market and do not respond to local renewable availability. This case considers a conventional hydrogen integration approach in which hydrogen is treated as an inflexible grid-connected load. It provides a reference for assessing the cost and system impacts of hydrogen production without local coordination or renewable access. As LEM and wind power are not available for electrolyzers, the values of $\mathrm{P}_{t,e}^{elz,lem}$ and $\mathrm{P}_{t,w,e}^{w\to elz}$ are set to zero.

*C. Case 2 – Market-Based Hydrogen Participation*

Case 2 allows electrolyzers to participate in the local electricity market while still prohibiting direct access to local wind generation. In this configuration, electrolyzers can purchase electricity either from the grid or through the local energy market based on pre-defined prices. Hydrogen production becomes price-responsive but remains indirectly coupled to renewable availability through market signals rather than physical resource sharing. Electrolyzers compete with prosumers for locally traded electricity, but wind generation continues to be allocated to households and storage before any grid export. This case represents a market-oriented hydrogen integration model commonly discussed in the literature, where flexibility is introduced through price signals alone, without explicit renewable allocation mechanisms. In this case, $\mathrm{P}_{t,w,e}^{w\to elz}$ is set to zero.

*D. Case 3 – Shared Wind Access*

Case 3 introduces direct competition between hydrogen production and prosumers for locally generated wind power. Electrolyzers are allowed to use wind energy directly, alongside households, subject to wind availability and system constraints. In this configuration, wind generation is jointly allocated among three possible uses: household consumption (including battery charging), hydrogen production, and grid export. This case describes a fully shared renewable access framework, where hydrogen acts as a flexible load that can absorb local renewable generation. It allows the study of trade-offs between hydrogen production and household flexibility resources when both compete for the same local renewable supply. This case is considered as a main case.

*E. Case 4 – Priority Wind Access*

Case 4 introduces a regulatory intervention in which hydrogen production is granted priority access to locally generated wind power. In this case, wind energy is first allocated to electrolyzers up to their operational demand, and only the remaining wind power is made available to prosumers and grid export. This priority is enforced through explicit allocation constraints rather than price signals. Hydrogen production is therefore excluded from competition with households during periods of high wind availability. This case is considered as a policy-driven case for hydrogen deployment. Such preferential renewable access or strategic prioritization of hydrogen production allows assessment of their system-level impacts. Priority access is implemented by introducing a wind allocation variable $\mathrm{P}_{t,e}^{w,pri}$ which represents wind power reserved for electrolyzers before allocation to other participants. This allocation is bounded by electrolyzer demand and total available wind generation.

$$\mathrm{P}_{t,e}^{w,pri} \leq \mathrm{P}_{t,e}^{elz} \tag{21}$$

$$\mathrm{P}_{t,e}^{w,pri} \leq \sum_{\mathrm{w}=1}^{\mathrm{W}} P_{t,w}^{wind} \tag{22}$$

These constraints ensure that priority allocation is feasible with respect to both electrolyzer demand and available wind generation. In the case when wind generation is less then electrolyzer demand, additional power is allowed to take from grid or LEM in same time period.

*F. Case 5 – Capped Wind Access*

Case 5 applies a constrained or moderated hydrogen participation mechanism, where electrolyzers are allowed to access local wind generation but only up to a predefined cap or share of 20%. Unlike Case 4, hydrogen does not receive full priority. Instead, its access to wind power is limited by explicit constraints that preserve a minimum share of renewable generation for prosumers. The cap can be interpreted as a regulatory safeguard designed to balance hydrogen deployment with household participation in the local market. This case represents a compromise between unrestricted competition (Case 3) and full prioritization (Case 4), assisting the evaluation of hybrid regulatory designs that aim to balance system cost reduction with fairness among market participants. This case is considered as a regulatory case. The case logic was implemented by following equation.

$$\sum_{e=1}^{\mathrm{E}} \sum_{\mathrm{w}=1}^{\mathrm{W}} \mathrm{P}_{\mathrm{t,w,e}}^{\mathrm{w\to elz}} \leq \alpha \sum_{\mathrm{w}=1}^{\mathrm{W}} \mathrm{P}_{\mathrm{t,w,e}}^{\mathrm{w\to elz}} \tag{23}$$

Where , $\alpha = 0.2$

Collectively, the six cases form a organized progression from no hydrogen integration to gradually coordinated and regulated hydrogen participation in local energy markets as shown in table 3. By comparing these cases under identical system conditions, the analysis isolates the role of market access rules in shaping renewable allocation, grid interaction, storage usage, and hydrogen production behavior. The resulting insights provide a basis for evaluating how different hydrogen integration strategies affect not only total system outcomes but also the distribution of benefits among local energy market participants.

TABLE III. CASE COMPARISON

| Case | $H_2$ | Wind allocation | | | Elz power from | | |
|---|---|---|---|---|---|---|---|
| | | E-U | Elz | Grid | Grid | LEM | Wind |
| 0 | × | ✓ | N.A. | ✓ | N.A. | N.A. | N.A. |
| 1 | ✓ | ✓ | × | ✓ | ✓ | × | × |
| 2 | ✓ | ✓ | × | ✓ | ✓ | ✓ | × |
| 3 | ✓ | ✓ | ✓ | ✓ | ✓ | ✓ | ✓ |
| 4 | ✓ | ✓ | ✓ | ✓ | ✓ | ✓ | ✓ |
| 5 | ✓ | ✓ | ✓ | ✓ | ✓ | ✓ | ✓ |

### G. Methodology

This section summarizes the optimization workflow. Hourly data for demand, generation, prices, storage, electrolyzers, and hydrogen demand are processed to define model inputs. Case-specific configuration determine hydrogen participation and renewable access rules. A step wise procedure is shown below.

**Algorithm 1. Procedure for LEM Optimization**

**Step 1:** Initialize Simulation: Set t ∈ $\mathcal{T}$, z ∈ $\mathcal{Z}$, w ∈ $\mathcal{W}$ and e ∈ $\mathcal{E}$

**Step 2:** Read Input Data: Import $P_{t,z}^{dem}$, $P_{t,z}^{pv}$, $P_{t,w}^{wind}$, electricity prices, battery parameters, electrolyzer parameters, and $D_t^{H2}$

**Step 3:** Pre-process Data: Convert all inputs into hourly resolution, construct local market prices, and prepare limits for grid, storage, and hydrogen systems.

**Step 4:** Select Case Configuration: Choose case (among {0,1,2,3,4,5}) and activate corresponding rules for hydrogen participation, wind access and market access.

**Step 5:** Define Decision Variables: Initialize variables for electricity trading (grid and LEM), battery operation, wind allocation, electrolyzer power, and hydrogen storage.

**Step 6:** Formulate System Constraints: Add power balance, LEM market clearing, battery dynamics, hydrogen balance, electrolyzer limits, wind allocation constraints.

**Step 7:** Apply Case-Specific Constraints: Modify the model based on selected case, including restrictions on wind-to-electrolyzer power, LEM participation, or capped/priority access.

**Step 8:** Define Objective Function: Minimize total system cost including grid purchases, LEM transactions, hydrogen production cost, and revenues from electricity export.

**Step 9:** Solve Optimization Problem: Solve the MILP model using CPLEX solver to obtain optimal schedules for all variables.

**Step 10:** Extract Results: Obtain electricity flows, battery operation, wind allocation, electrolyzer operation, and hydrogen production profiles.

**Step 11:** Compute Performance Metrics: Calculate total cost, energy volumes, renewable utilization, battery usage, and hydrogen production indicators.

**Step 12:** Store and Visualize Outputs: Save results and generate time-series plots and comparative analysis across cases.

**Step 13:** End Procedure: Finalize simulation and prepare outputs for further analysis.

## VI. RESULTS & DISCUSSION

The key results are shown in table 4. In the base case without hydrogen, the system operates with low cost and stable behavior. For example, the total cost in W1 is about 366,545 NOK, and grid imports remain moderate (around 259 MWh for a total load of 306.5 MWh). Batteries handle most of the flexibility by shifting energy across time. When hydrogen is introduced as a grid-only load, system cost increases significantly. In W1, the cost rises to about 675,881 NOK, which is an increase of nearly 85% compared to the base case. This increase is mainly due to hydrogen production relying entirely on grid electricity, with hydrogen power cost alone exceeding 300,000 NOK. Allowing hydrogen to participate in the local market improves performance. In Case 2, the total cost drops to about 586,813 NOK in W1, which is almost 13% lower than the grid-only case. Grid imports are also reduced significantly (from about 259 MWh to 177 MWh), as hydrogen production shifts to lower-price periods. However, wind is still not directly used by hydrogen, and its allocation to households remains unchanged.

Results are also interesting when hydrogen is allowed to access wind generation. In the shared access case, hydrogen begins to use a portion of wind energy, reaching about 16 MWh in W3 and 22 MWh in W4. This reduces the amount of wind available to households and slightly lowers system cost compared to Case 2. When priority access is introduced, the effect becomes much stronger. Hydrogen receives a large share of wind energy; about 32 MWh in W1 and over 55 MWh in W4. As a result, hydrogen power cost decreases (around 6–10% lower than Case 3), and total system cost also drops slightly. However, wind supplied to households decreases sharply (for example from about 48 MWh to 16 MWh in W1), which affects battery charging and increases reliance on grid electricity in some periods. The capped access, case 5 provides a more balanced outcome. Hydrogen receives a limited share of wind (8–11 MWh), while households retain most of the renewable supply.

Fig. 4 shows the time-series heat map of output variables of the system under Case 3 (shared wind access) across representative weeks. The electrolyzer does not operate continuously but is activated during periods of higher wind availability and lower local prices. The hydrogen storage profiles exhibit a cyclic pattern, with storage charging during high production periods and discharging to meet demand over time. Overall, the figure shows that under shared access, hydrogen operates as a flexible load that responds to both market signals and local renewable conditions.

TABLE IV. KEY RESULTS

| | Objective function (NOK) | | | | H2 power cost (NOK) | | | |
|---|---|---|---|---|---|---|---|---|
| | W1 | W2 | W3 | W4 | W1 | W2 | W3 | W4 |
| Case 0 | 366544.84 | 63247.97 | 18630.77 | 34653.89 | N.A. | N.A. | N.A. | N.A. |
| Case 1 | 675881.00 | 204106.56 | 196203.50 | 195636.00 | 309336.23 | 140837.09 | 177469.32 | 160983.87 |
| Case 2 | 586813.28 | 132379.45 | 151014.94 | 138584.07 | 271012.14 | 105704.03 | 144547.95 | 122975.77 |
| Case 3 | 586604.92 | 132234.28 | 148389.96 | 133168.26 | 270918.97 | 105477.02 | 141543.64 | 117216.23 |
| Case 4 | 582536.62 | 126411.63 | 136076.61 | 114763.42 | 255165.20 | 99572.52 | 130483.98 | 100628.99 |
| Case 5 | 585779.14 | 130516.03 | 147865.25 | 133780.58 | 267018.31 | 103991.99 | 141498.32 | 118327.18 |
| | Energy import from grid by end users (kWh) | | | | Energy import from LEM by end users (kWh) | | | |
| Case 0 | 259206.40 | 130417.28 | 39574.99 | 31308.33 | 4102.54 | 59468.48 | 69695.06 | 53793.58 |
| Case 1 | 259206.11 | 130287.49 | 38690.43 | 31414.75 | 4356.03 | 59286.46 | 69037.67 | 54322.96 |
| Case 2 | 176682.89 | 71968.19 | 27083.40 | 3322.17 | 138526.05 | 148093.03 | 128176.38 | 117006.33 |
| Case 3 | 176433.24 | 72395.55 | 26902.57 | 3385.23 | 139239.61 | 148432.06 | 120732.93 | 108405.63 |
| Case 4 | 201965.61 | 75513.06 | 26788.07 | 2862.70 | 120842.16 | 149550.68 | 131225.20 | 117407.06 |
| Case 5 | 182552.70 | 72373.51 | 27062.28 | 3232.52 | 139608.08 | 147898.43 | 127009.88 | 115293.01 |
| | Wind to Electrolyzer energy (kWh) | | | | Wind to end user energy (kWh) | | | |
| Case 0 | N.A. | N.A. | N.A. | N.A. | 48331.72 | 35933.55 | 47420.99 | 56315.67 |
| Case 1 | N.A. | N.A. | N.A. | N.A. | 48331.72 | 35933.55 | 47420.99 | 56315.67 |
| Case 2 | N.A. | N.A. | N.A. | N.A. | 48331.72 | 35933.55 | 47420.99 | 56315.67 |
| Case 3 | 0.0 | 1759.27 | 15991.19 | 22637.89 | 48331.72 | 34174.28 | 31429.80 | 33677.78 |
| Case 4 | 31786.51 | 18929.67 | 38855.55 | 55080.81 | 16545.20 | 17003.87 | 8565.44 | 1234.86 |
| Case 5 | 7854.47 | 4821.20 | 8183.52 | 10941.08 | 40477.24 | 31112.35 | 40317.81 | 45374.59 |

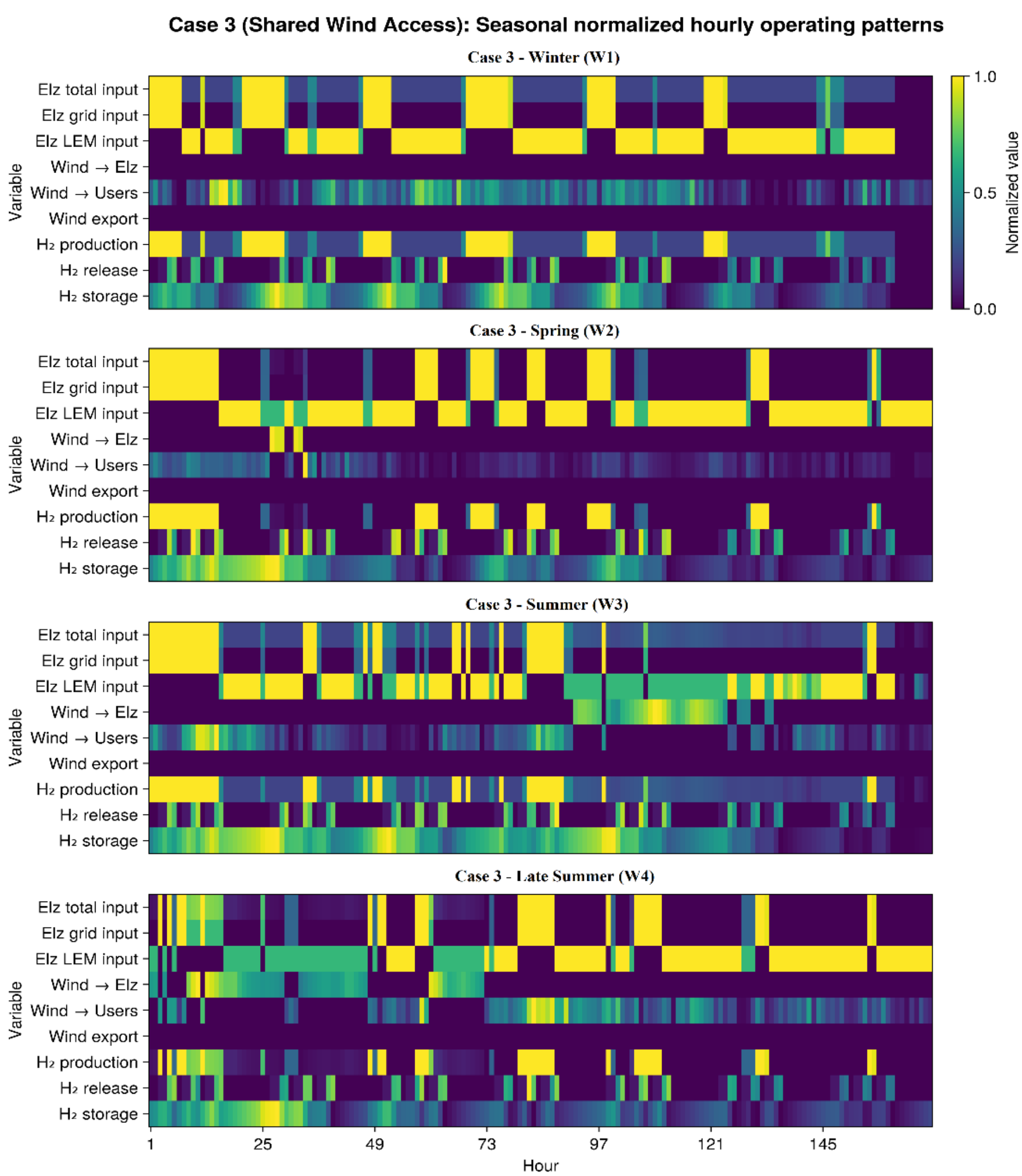

Fig. 4. Main case performance

Seasonal variation also affects the results. In weeks with higher wind availability, especially W3 and W4, hydrogen absorbs more renewable energy when access is allowed, leading to better utilization of local resources. In contrast, during lower wind periods, hydrogen depends more on grid electricity, and the impact of access rules becomes smaller. Across all cases, it is evident that market rules clearly affect system behavior. Price-based participation improves operation compared to grid-only hydrogen, but it does not change how renewables are allocated. Direct access to wind enables hydrogen to act as a flexible load, but it also creates competition with households. The results therefore show that hydrogen integration cannot be evaluated only in terms of cost or efficiency. The way renewable energy is shared between participants has a direct impact on system performance and on how benefits are distributed within the local market.

## VII. Conclusion

The results show that hydrogen integration in local energy systems is affected by the rules that govern access to renewable resources. When electrolyzers are allowed to respond to local renewable availability, hydrogen production aligns with periods of high wind generation and reduces reliance on grid electricity. However, price-based participation alone does not ensure that hydrogen production is driven by renewable energy. Shared access allows hydrogen and households to compete for wind power, while priority access shifts a larger portion of renewable energy toward hydrogen production. The results also reveal interactions between hydrogen production and household battery storage. Priority access improves hydrogen operation but reduces renewable availability for prosumers, whereas capped access mechanisms moderate this effect and provide a more balanced allocation of local resources.

These findings indicate that regulatory rules can play an important role in design of future electricity–hydrogen market designs. The results indicate that price signals alone may not be sufficient for hydrogen integration. Overall, hydrogen should be viewed as an active market participant rather than a conventional electrical load. By comparing alternative renewable access mechanisms within a unified optimization framework, the paper contributes to the understanding of how hydrogen can be integrated into community energy systems while maintaining system flexibility and participant interests.